\documentclass[aps, pra, superscriptaddress, showpacs, twocolumn, nofootinbib]{revtex4-1}

\usepackage{amsfonts, amsmath, amssymb, graphicx, hyperref}

\begin{document}

\title{Tensor network states in time-bin quantum optics}

\author{Michael Lubasch}
\affiliation{Clarendon Laboratory, Department of Physics, University of Oxford, Oxford OX1 3PU, United Kingdom}

\author{Antonio A. Valido}
\affiliation{QOLS, Blackett Laboratory, Imperial College London, London SW7 2AZ, United Kingdom}

\author{Jelmer J. Renema}
\affiliation{Clarendon Laboratory, Department of Physics, University of Oxford, Oxford OX1 3PU, United Kingdom}

\author{W. Steven Kolthammer}
\affiliation{QOLS, Blackett Laboratory, Imperial College London, London SW7 2AZ, United Kingdom}

\author{Dieter Jaksch}
\affiliation{Clarendon Laboratory, Department of Physics, University of Oxford, Oxford OX1 3PU, United Kingdom}
\affiliation{Center for Quantum Technologies, National University of Singapore, 117543 Singapore}

\author{M. S. Kim}
\affiliation{QOLS, Blackett Laboratory, Imperial College London, London SW7 2AZ, United Kingdom}

\author{Ian Walmsley}
\affiliation{Clarendon Laboratory, Department of Physics, University of Oxford, Oxford OX1 3PU, United Kingdom}

\author{Ra\'{u}l Garc\'{i}a-Patr\'{o}n}
\affiliation{Centre for Quantum Information and Communication, Ecole Polytechnique de Bruxelles, CP 165, Universit\'{e} Libre de Bruxelles, 1050 Brussels, Belgium}


\begin{abstract}
The current shift in the quantum optics community towards experiments with many modes and photons necessitates new classical simulation techniques that efficiently encode many-body quantum correlations and go beyond the usual phase space formulation.
To address this pressing demand we formulate linear quantum optics in the language of tensor network states.
We extensively analyze the quantum and classical correlations of time-bin interference in a single fiber loop.
We then generalize our results to more complex time-bin quantum setups and identify different classes of architectures for high-complexity and low-overhead boson sampling experiments.
\end{abstract}

\date{\today}

\maketitle

\section{Introduction}

Because the realization of a universal quantum computer via quantum optical elements is a challenging task, the quantum optics community has recently turned its attention to non-universal quantum devices.
These quantum devices require fewer and simpler components but nevertheless have the potential to demonstrate a quantum advantage over classical computing devices~\cite{HaMo17}.
A famous example is the boson sampling proposal by Aaronson and Arkhipov in 2011~\cite{AaAr11} about a quantum device in which many photons interfere in a planar interferometer and are counted at the output.
Aaronson and Arkhipov showed that classical computers cannot efficiently simulate large boson samplers.
This has led to a significantly increased interest in multi-photon interference experiments.

The new territory between the quantum advantage regime of boson sampling and earlier few-photon experiments offers new possibilities for scientific discoveries and perspectives on quantum optics, but it also poses a new challenge.
Traditional mathematical tools from quantum optics are not suitable anymore for the theoretical analysis of this new territory.
E.g.\ the phase space representation of quantum states~\cite{Pe86} has proven to be extremely useful for simulating the evolution of multi-mode Gaussian states under Gaussian operations and measurements~\cite{BaSa02, MaEi12}.
These techniques have also been extended to non-Gaussian mixtures of pure Gaussian states~\cite{VeWiFeEm13} and can be applied to non-Gaussian states with sufficiently large noise~\cite{RaRaCa16}.
However, these methods get costly for superpositions of many photon numbers, photon-subtracted Gaussian states, multi-mode entangled non-Gaussian states, and even already for the sampling of a Gaussian state in its photon number basis.
Thus there exists a need for new theoretical tools that can tackle the large multi-photon interference experiments that we are witnessing now.
We believe that a promising new approach is given by tensor network states, which represent a powerful theoretical framework to efficiently describe highly-correlated quantum states and operations~\cite{VeMuCi08, CiVe09, Ei13, Or14}.

\begin{figure}
\centering
\includegraphics[width=0.9\linewidth]{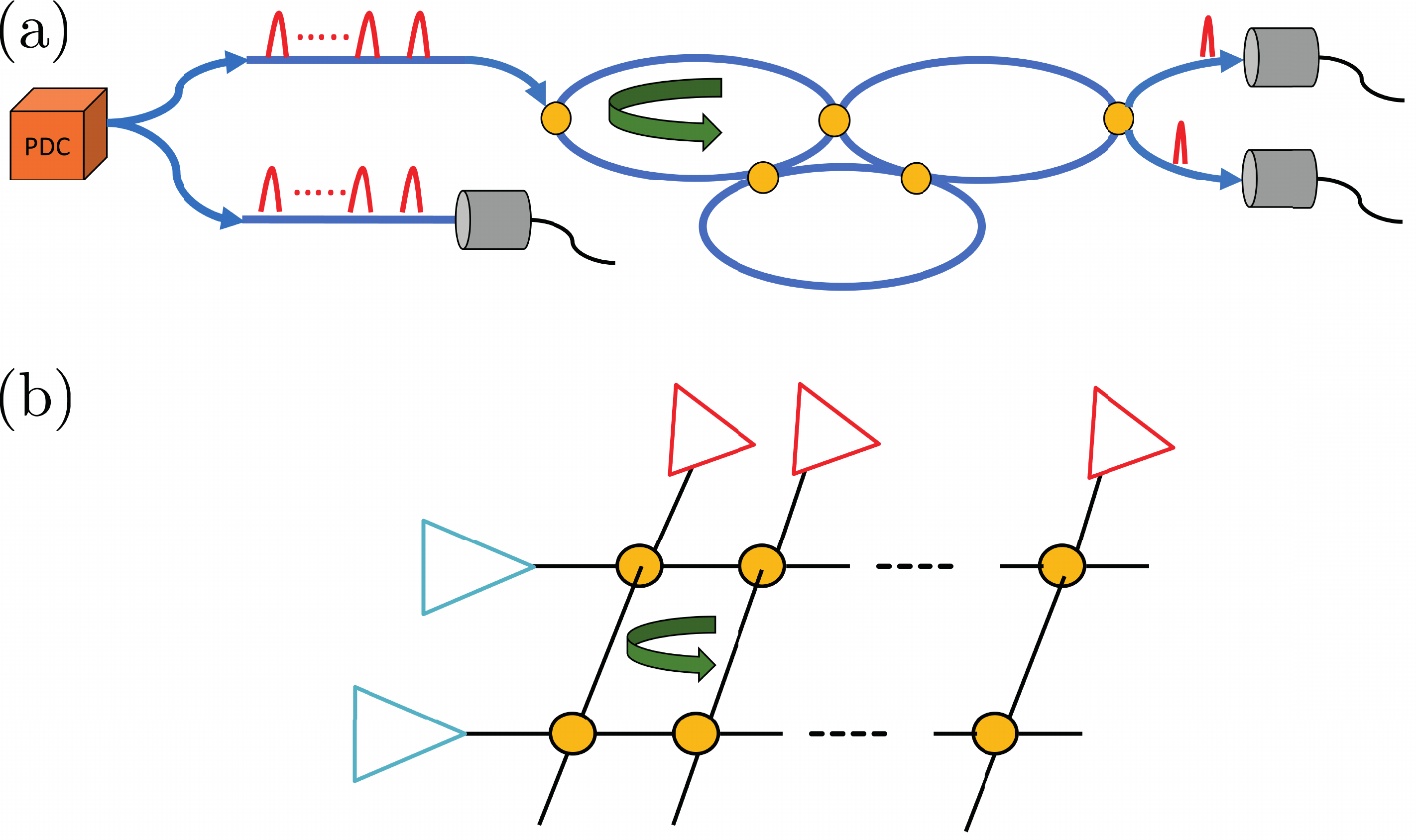}
\caption{\label{fig:1}
(Color online.)
(a) A quantum optical interferometer is composed of: a heralded photon source, i.e.\ a parametric-down-conversion source (orange box $PDC$) generating pairs of photons that are heralded by photon-counting detectors on the ancillary mode; a series of interferometric loops (green arrow) interconnected by couplers (yellow circles) mediating the coupling between optical loops; a set of outputs that are photo-counted.
(b) Using the techniques described in this article one can build the tensor network state corresponding to an optical interferometer.
Every node in the graph is a tensor representing either an input state (red triangle at the top for a time-bin input state and blue triangle at the left for an initial loop state), a coupling or local phase operation (yellow circle), or a measurement outcome.
The connecting edges are causal connections between those elements and correspond to tensor contractions.
Open edges represent the set of outputs and take values in the photon number basis corresponding to the possible measurement outcomes.
}
\end{figure}

Here we adapt the formalism of tensor network states to model photonic quantum interference in time-bin quantum optics~\cite{ScEtAl10}.
In this context the quantum information is encoded in the temporal degrees of freedom, i.e.\ arrival times, of photons rather than in their spatial degrees of freedom.
This has the advantage that the same sources and detectors can be repeatedly used in the same experiment.
Various experimental techniques have been proposed to implement mode mixing between adjacent temporal modes, e.g.\ based on orbital angular momentum~\cite{GoEtAl13, CaEtAl15} or dispersion~\cite{PaEn16}, but the most common technique is temporal delays, which are usually experimentally realized by a network of fiber loops~\cite{ScEtAl10}.
Fig.~\ref{fig:1} (a) shows an example.
Such networks of fiber loops have been used for boson sampling~\cite{MoEtAl14, HeEtAl17} as well as for the realization of multidimensional quantum walks~\cite{ScEtAl12}, large-scale quantum walks~\cite{BoEtAl16}, and continuous-variable quantum information processing~\cite{TaFu17}.

The complexity\footnote{
In this article we use the notion of complexity in a general sense to describe systems or problems that are hard to simulate or solve.
We use the notion of computational cost to denote upper bounds on the scaling of the running time and memory requirements of an algorithm.
And we use the notion of computational complexity or hardness for results that were obtained via rigorous proofs using complexity theory.}
of the information processing which such a network can implement depends on the nature of the coupling between the adjacent time-bin modes.
If we allow the coupling between adjacent modes to change with each time step, then it is known that an arbitrary interferometer can be implemented~\cite{MoEtAl14}, and hence the hardness of sampling from the output state of such a system follows from Ref.~\cite{AaAr11}.
However, such control might not always be achievable, especially if low-loss operation is desired.
Fast modulation requires the use of the electro-optic effect, and it has proven difficult to implement a low-loss modulator based on this effect due to the technical difficulty of realizing low-loss lithium niobate waveguides~\cite{WaEtAl14, WoEtAl00, Kr17}.
Therefore it is interesting to ask what degree of complexity can be achieved by static couplers, i.e.\ couplers that have a constant value throughout the experiment.

In this article, we analyze the complexity of such static time-bin experiments from the perspective of tensor network states -- also known as tensor networks.
Tensor networks have successfully been utilized in approximately solving high-dimensional optimization problems such as finding ground states of strongly correlated quantum many-body systems~\cite{Wh92, OeRo95, VePoCi04, Sc05, Sc11} (and nowadays their applicability goes far beyond that: see e.g.~Ref.~\cite{LuFuApRuCiBa16} for a tensor network application in density functional theory, or Ref.~\cite{LuMoJa18} for an application in the context of general partial differential equations).
In the tensor network language, the quantum state or operation is represented by a graph where nodes correspond to tensors and edges indicate tensor contraction rules -- an example is shown in Fig.~\ref{fig:1} (b).
The central numerical ingredient in all tensor network computations is the contraction of possibly large products of tensors.
The computational cost of these tensor product contractions increases e.g.\ with the number of closed loops contained in the tensor network graph.
We show in this work that there exists a connection between loops in a quantum optical interferometer and loops in the corresponding tensor network description.
We use this connection to study the computational complexity of a boson sampling experiment.

Our results consist of three parts.
First, we formalize the connection between time-bin interferometers and tensor networks.
Building on previous works~\cite{IbOrLa07, OhKiEi10, OhEi12, TeWo12, PiZo16, VeGuPiZo17, PiChZoLu17}, we show that a quantum optical setup can be modeled by a tensor network.
Second, we consider the states generated by the simplest possible time-bin system: a single fiber loop.
In the tensor network formalism this is represented by a matrix product state~\cite{FaNaWe92, OeRo95, Vi03, PeVeWoCi07}.
Matrix product states are part of the set of numerically tractable tensor networks which can prove useful for many interesting quantum optical applications such as e.g.\ metrology or sensing~\cite{JaDe13}.
Finally, from known conditions for tensor networks to be efficiently contractible on classical computers, we discuss experimentally feasible designs of time-bin interferometers that could provide a quantum advantage.

\section{Single fiber loop}
\label{sec:fiberloop}

\begin{figure}
\centering
\includegraphics[width=0.9\linewidth]{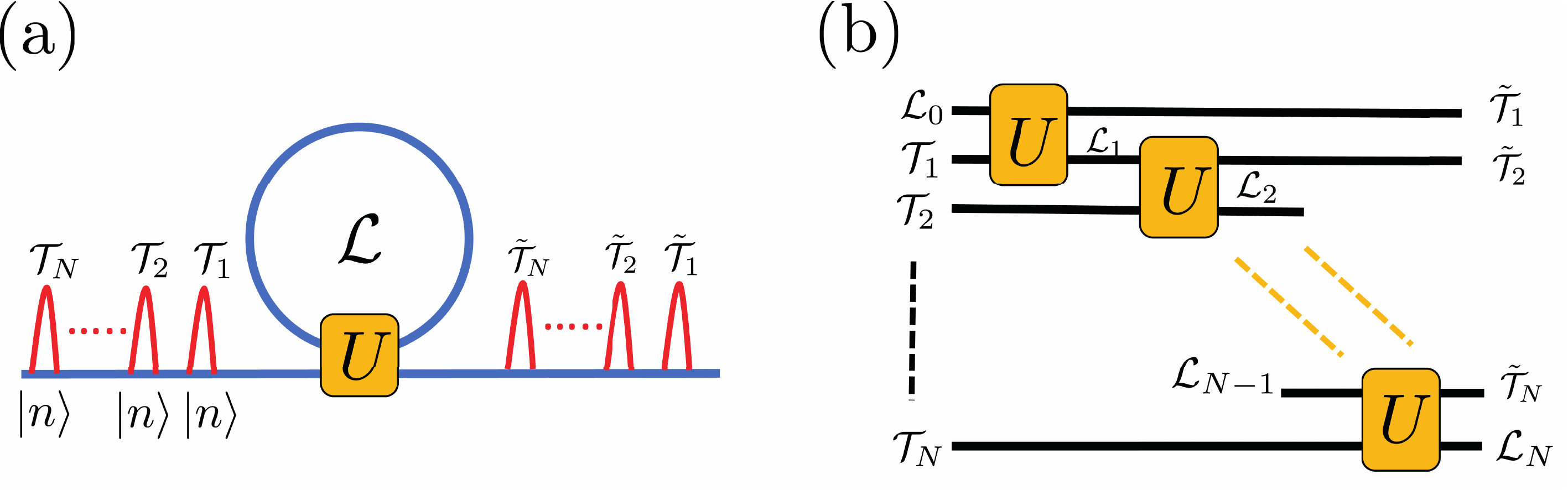}
\caption{\label{fig:2}
(Color online.)
(a) A single fiber loop with internal degree of freedom $\mathcal{L}$ interacts with $N$ input time-bin modes $\mathcal{T}_{i}$ through a time-invariant coupler implementing the unitary $U$ given in Eq.~\eqref{eq:BSE}.
This process has $N$ output time-bin modes $\tilde{\mathcal{T}}_{i}$.
(b) A circuit representation unfolding the fiber loop coupling with the $N$ time-bin modes.
The circuit is composed of a ladder of identical coupler gates realizing the unitary $U$ between the time-bin mode $\mathcal{T}_{i}$ and the fiber loop mode $\mathcal{L}_{i}$ at time $i$.
Note that in the circuit representation the fiber loop system $\mathcal{L}$ moves one step down the ladder after every coupling.
}
\end{figure}

The central topic of this article is time-bin quantum optical interferometry which has as its most basic element the single fibre loop, shown in Fig.~\ref{fig:2} along with its representation as a planar quantum circuit.
This system is composed of a single time delay $\mathcal{L}$ -- which is the fiber loop -- and a pulse train on $N$ time-bin modes $\mathcal{T}_{i}$ that are equally spaced and sequentially interact with the fiber loop $\mathcal{L}$ through a coupler at time steps $\Delta t$.
Such a system is experimentally realized by setting the length of the loop equal to the distance light travels between two input time-bins, in order to achieve temporal overlap between photons entering the loop at time step $i$ and photons arriving at time step $i+1$.
The interference between modes $i$ and $i+1$ is mediated by a time-independent coupler that implements a unitary $U : \mathcal{H}_{\mathcal{T}_{i}} \otimes \mathcal{H}_{\mathcal{L}_{i}} \rightarrow \mathcal{H}_{\tilde{\mathcal{T}}_{i}} \otimes \mathcal{H}_{\tilde{\mathcal{L}}_{i}}$, where we have used the tilde in $\tilde{\mathcal{T}}_{i}$ and $\tilde{\mathcal{L}}_{i}$ to denote the output nature of that mode at time step $i$.
The coupling is the well-known passive bosonic hopping coupling~\cite{CaSaTe89, KiSoBuKn02} equivalent to a beam splitter:
\begin{eqnarray}\label{eq:BSE}
\hat{U}(\theta, \phi) & = & \mathrm{e}^{\theta \left( \hat{a}^{\dagger} \hat{b} \mathrm{e}^{\mathrm{i} \phi} - \hat{a} \hat{b}^{\dagger} \mathrm{e}^{-\mathrm{i} \phi} \right) } ,
\end{eqnarray}
where $\hat{a}$ and $\hat{b}$ are the loop and time-bin annihilation operators, $\phi$ represents a phase between the output modes, and $\theta$ parametrizes the intensity transmission and reflection probabilities $\cos^{2}(\theta)$ and $\sin^{2}(\theta)$, respectively.
In this context, transmission and reflection refer to the processes of photons staying in the same mode and photons moving to the other mode, respectively.
Note that $\hat{a}$, $\hat{b}$, and $\hat{U}$ have an index $i$ that we leave out in Eq.~\eqref{eq:BSE} for notational simplicity.
This system will be explained and analyzed in detail in Sec.~\ref{sec:classicalsimu}.
There we will also present some subtleties that appear in tensor network simulations of quantum optics.

\section{Matrix product states}
\label{sec:TN}

\begin{figure}
\centering
\includegraphics[width=0.9\linewidth]{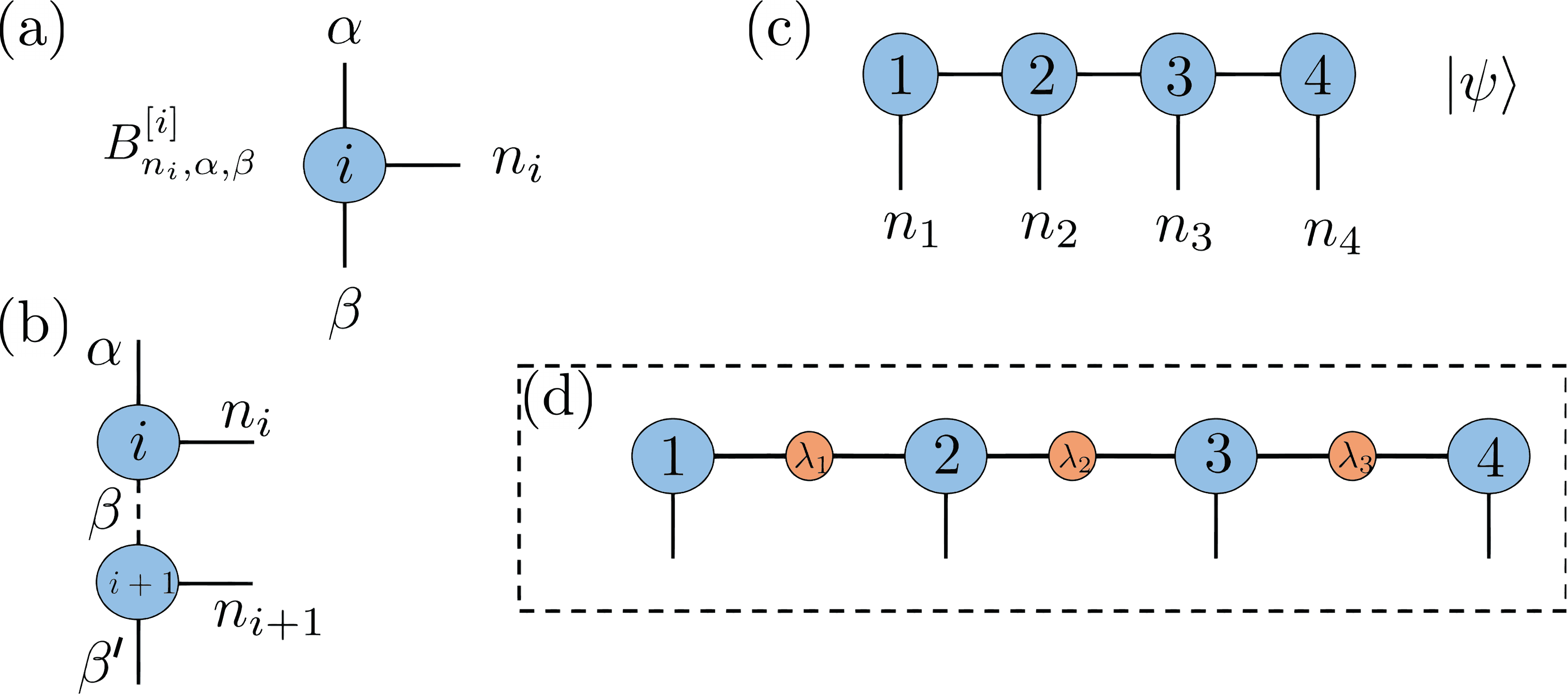}
\caption{\label{fig:3}
(Color online.)
(a) An example of a tensor $B^{[i]}_{n_{i}, \alpha, \beta}$ of rank three represented by a vertex $i$ with three edges for the three indices $n_{i}$, $\alpha$, and $\beta$.
For the quantum optical problems discussed in this article, the index $n_{i}$ represents a physical quantity -- namely the number of photons in time-bin mode $\mathcal{T}_{i}$ -- and the indices $\alpha$ and $\beta$ represent virtual quantities -- encoding the correlations between adjacent time-bin modes.
(b) The tensor contraction of $B^{[i]}$ and $B^{[i+1]}$ along their shared index (dashed line) is an operation equivalent to matrix multiplication along the shared index, as explained in the main text.
(c) Graphical representation of a quantum state $|\psi\rangle$ corresponding to a matrix product state of four tensors, as defined in Eq.~\eqref{eq:MPSd}.
(d) All matrix product states can be transformed into a canonical form where a diagonal matrix $\lambda$ of Schmidt coefficients is assigned to every edge of the matrix product state, see Eq.~\eqref{eq:canonical} and corresponding discussion.
}
\end{figure}

The central ingredient of all tensor network calculations is the tensor contraction, i.e.\ a generalization of vector and matrix multiplication to tensors of arbitrary rank.
An example is shown in Fig.~\ref{fig:3}.
The contraction of two rank-three tensors $B^{[i]}$ and $B^{[i+1]}$ along their shared index gives a rank-four tensor defined as $T_{\alpha, n_{i}, n_{i+1}, \beta^{'}} = \sum_{\beta} B^{[i]}_{\alpha, n_i, \beta} B^{[i+1]}_{\beta, n_{i+1}, \beta^{'}}$.
One of the most widely used tensor network states is the matrix product state~\cite{FaNaWe92, OeRo95, Vi03, PeVeWoCi07} shown in Fig.~\ref{fig:3} (c).
This is a one-dimensional tensor network that can e.g. represent a quantum state $|\psi\rangle$ of a quantum system composed of $N$ $d$-dimensional subsystems.
We denote the computational basis states of the individual subsystems $i$ by $|n_{i}\rangle$ where $n_{i} \in \{ 0, 1, \ldots, d-1 \}$.
Then a matrix product state takes on the form
\begin{eqnarray}\label{eq:MPSd}
|\psi\rangle & = & \sum_{n_{1}, \ldots, n_{N}} \left( B^{[1]}_{n_{1}} B^{[2]}_{n_{2}} \ldots B^{[N]}_{n_{N}} \right) | n_{1}, \ldots, n_{N} \rangle ,
\end{eqnarray}
where the sum $\sum_{n_{1}, n_{2}, \ldots, n_{N}} = \sum_{n_{1}=0}^{d-1} \sum_{n_{2}=0}^{d-1} \ldots \sum_{n_{N}=0}^{d-1}$ runs over all basis states $| n_{1}, n_{2}, \ldots, n_{N} \rangle$.
Here, for each value of $n_{1}$, $B^{[1]}_{n_{1}}$ is a transposed vector of dimension $\chi_{1}$, for each value of $n_{N}$, $B^{[N]}_{n_{N}}$ is a vector of dimension $\chi_{N}$, and for each value of $n_{i}$ where $1 < i < N$, $B^{[i]}_{n_{i}}$ is a matrix of dimension $\chi_{i-1} \times \chi_{i}$.
Thus in a matrix product state the wave function coefficient for a specific basis state $ | n_{1}, n_{2}, \ldots, n_{N} \rangle$ is given by a vector-matrix-matrix-\ldots-matrix-vector product $B^{[1]}_{n_{1}} B^{[2]}_{n_{2}} B^{[3]}_{n_{3}} \ldots B^{[N-1]}_{n_{N-1}} B^{[N]}_{n_{N}}$.

The dimensions $\chi_{i}$ determine the computational cost of storing and processing matrix product states.
We define $\chi = \max_{i}(\chi_{i})$ and call this the bond dimension.
Then the total number of parameters and thus the computational cost of storage for such a matrix product state scales as $\mathcal{O}(Nd\chi^{2})$.
The computational cost of processing such a matrix product state depends on the details of the algorithm.
However, most matrix product state algorithms have a computational cost scaling like $\mathcal{O}(Nd^{2}\chi^{2})+\mathcal{O}(Nd\chi^{3})$~\cite{VeMuCi08, Or14}.

The bond dimension also determines the amount of entanglement that a matrix product state can contain.
Every matrix product state can be transformed into a canonical form~\cite{Vi03, PeVeWoCi07}
\begin{eqnarray}\label{eq:canonical}
|\psi\rangle & = & \sum_{n_{1}, \ldots, n_{N}} \left( \Gamma^{[1]}_{n_{1}} \lambda^{[1]} \Gamma^{[2]}_{n_{2}} \lambda^{[2]} \ldots \Gamma^{[N]}_{n_{N}} \right) | n_{1}, \ldots, n_{N} \rangle , \quad
\end{eqnarray}
in which the $\lambda^{[i]}$ are diagonal matrices containing the Schmidt coefficients for the matrix product state bipartition between sites $1, 2, \ldots, i$ and $i+1, i+2, \ldots, N$ (see Ref.~\cite{NiCh10} for details).
A graphical representation of a matrix product state in canonical form is sketched in Fig.~\ref{fig:3} (d).
The canonical form allows us to compute the entanglement entropy~\cite{BeBePoSc96, HoHoHoHo09} for every bipartition of the matrix product state as
\begin{eqnarray}\label{eq:E}
E(i) & = & -\sum_{k=0}^{\chi_{i}-1} (\lambda^{[i]}_{k})^{2} \log_{2} \left( (\lambda^{[i]}_{k})^{2} \right) .
\end{eqnarray}

The entanglement entropy plays an important role in tensor network simulations of many-body systems.
Eq.~\eqref{eq:E} tells us that a matrix product state of bond dimension $\chi$ can never have values of $E$ larger than $\log_{2}(\chi)$.
Therefore, when a many-body system contains large amounts of entanglement, then the matrix product state simulation requires large values of $\chi$ and thus has a high computational cost.
Thus entanglement can be used as an indicator for the hardness of the classical simulation based on matrix product states~\cite{ScWoVeCi08}.

Tensor network techniques can be applied to quantum optics in different ways~\cite{IbOrLa07, OhKiEi10, OhEi12, TeWo12}.
In this article we map time-bin quantum optical setups to their corresponding tensor network state by identifying the time-bin modes as the individual quantum systems in the tensor network description.
Then each time-bin mode $\mathcal{T}_{i}$ is described by a $d$-dimensional Hilbert space and bosonic mode occupation numbers $n_{i} \in \{ 0, 1, \ldots, d-1 \}$.
Therefore our tensor network description implies a maximally possible bosonic mode occupation number $n_{\mathrm{max}} = d-1$ for each time-bin mode.
At first glance, this seems to be a strong restriction for a general quantum optical setup in which e.g.\ photon bunching might occur and thus a large bosonic mode occupation number might appear in a single time-bin mode.
However, we will see in the next Sec.~\ref{sec:classicalsimu} that this is not a problem for tensor network algorithms.
The parameters $d$ and $\chi$ are always chosen large enough to capture the true physics.
This is achieved by successively increasing $d$ and $\chi$ until all results do not change anymore and have thus become independent of $d$ and $\chi$.

\section{Efficient classical simulation of a single fiber loop}
\label{sec:classicalsimu}

In this section we illustrate how tensor networks can be used in quantum optics to analyze the coupling of $N$ time-bin modes over a single fiber loop.
Such a system takes on the form of a matrix product state.
Using this matrix product state description, we show how entanglement and correlations are computed and explain how boson sampling can be done efficiently on a classical computer.

\subsection{Matrix product states from a single fiber loop}
\label{sec:MPSanalysis}

\begin{figure}
\centering
\includegraphics[width=0.6\linewidth]{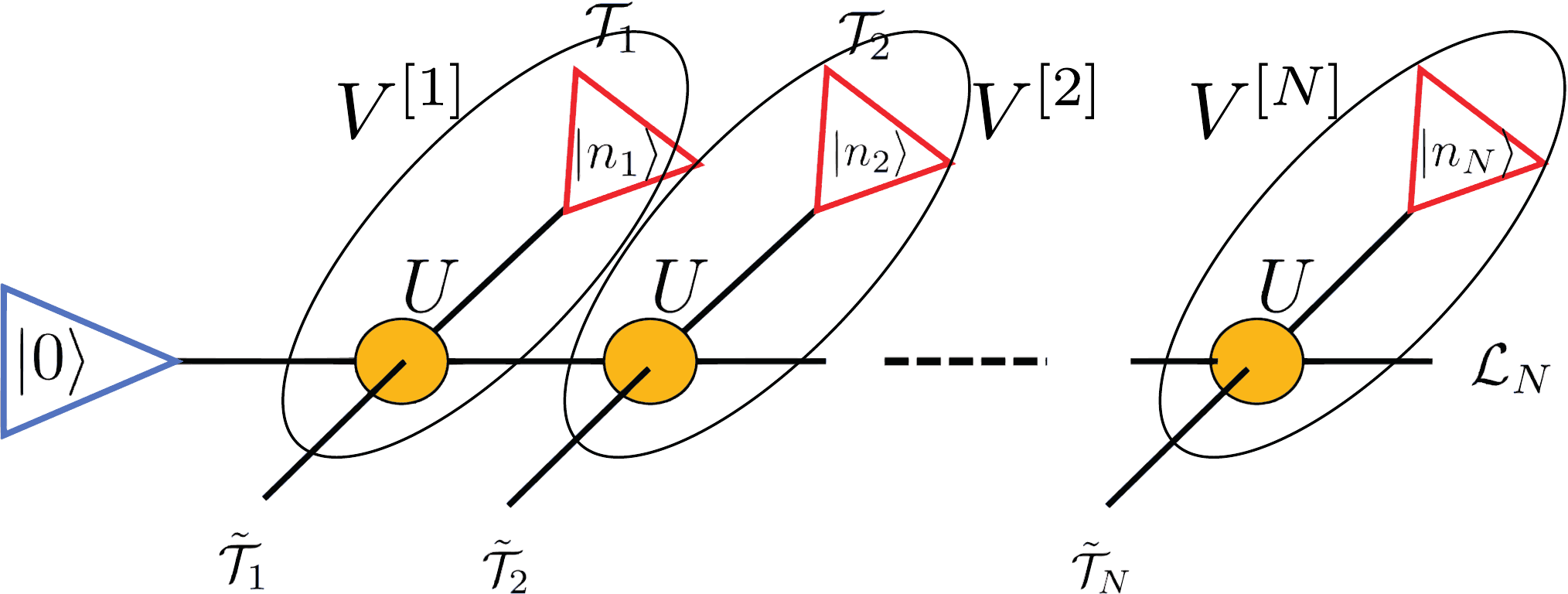}
\caption{\label{fig:4}
(Color online.)
Tensor network representation of the output state of the $N$ time-bin modes and the loop at step $N$.
Each yellow circle corresponds to a tensor with two input and two output indices where each index represents either an input or output degree of freedom of the coupling $U$.
The blue (left) and red (top) triangles correspond to vectors, i.e.\ tensors of rank one, that represent the input states of the loop and time-bins, respectively.
}
\end{figure}

The single fiber loop architecture corresponds to a sequential coupling between an ancillary system, represented here by the fiber loop $\mathcal{L}$, and a train of input systems, represented here by the time-bin modes $\mathcal{T}_{i}$.
We model the sequential process by a matrix product state~\cite{ScSoVeCiWo05, ScHaWoCiSo07}, see also Fig.~\ref{fig:4}.

For a train of $N$ input time-bin modes $\mathcal{T}_{i}$ in the multi-mode Fock state $|n_{1}\rangle \otimes |n_{2}\rangle \otimes \ldots \otimes |n_{N}\rangle$ the final state reads
\begin{eqnarray}\label{eq:TMFLS}
|\psi\rangle & = & V^{[N]} V^{[N-1]} \ldots V^{[2]} V^{[1]} |\varphi\rangle_{\mathcal{L}} .
\end{eqnarray}
Here $|\varphi\rangle_{\mathcal{L}}$ denotes the initial state of the loop which we always set to be the vacuum state $|0\rangle$ and $V^{[i]}$ denotes the isometry obtained from $U$ of Eq.~\eqref{eq:BSE} after action of the time-bin input state $|n_{i}\rangle$:
\begin{eqnarray}\label{eq:EqV1}
V^{[i]} & = & \hat{U} |n_{i}\rangle_{\mathcal{T}_{i}}\\
          & = & \left( \sum_{o, p, q, r} U_{o, p}^{q, r} |o\rangle_{\tilde{\mathcal{L}}_{i}} |p\rangle_{\tilde{\mathcal{T}}_{i}} \langle q|_{\mathcal{L}_{i}} \langle r|_{\mathcal{T}_{i}} \right) |n_{i}\rangle_{\mathcal{T}_{i}}\\
          & = & \sum_{o, p, q} U_{o, p}^{q, n_{i}} |o\rangle_{\tilde{\mathcal{L}}_{i}} |p\rangle_{\tilde{\mathcal{T}}_{i}} \langle q|_{\mathcal{L}_{i}} ,
\end{eqnarray}
where $\mathcal{L}_{i}$/$\tilde{\mathcal{L}}_{i}$ denotes the loop input/output and $\mathcal{T}_{i}$/$\tilde{\mathcal{T}}_{i}$ denotes the time-bin input/output mode of $U$ at time step $i$.
If we write all time-bin output states $|\tilde{n}_{i}\rangle$ and the final loop output state $|\tilde{n}_{\tilde{\mathcal{L}}}\rangle$ explicitly then Eq.~\eqref{eq:TMFLS} turns into
\begin{widetext}
\begin{eqnarray}\label{eq:FS1}
|\psi\rangle & = & \sum_{\tilde{n}_{1}, \tilde{n}_{2}, \ldots, \tilde{n}_{N}, \tilde{n}_{\tilde{\mathcal{L}}}} V^{[N]}_{\tilde{n}_{\tilde{\mathcal{L}}}, \tilde{n}_{N}} V^{[N-1]}_{\tilde{n}_{N-1}} \ldots V^{[2]}_{\tilde{n}_{2}} V^{[1]}_{\tilde{n}_{1}} |\varphi\rangle_{\mathcal{L}} |\tilde{n}_{\tilde{\mathcal{L}}}, \tilde{n}_{N}, \ldots, \tilde{n}_{2}, \tilde{n}_{1} \rangle
\end{eqnarray}
\end{widetext}
where $V^{[1]}$ acts on $|\varphi\rangle_{\mathcal{L}}=|0\rangle$ and $|\tilde{n}_{\tilde{\mathcal{L}}}\rangle$ represents the loop output mode of $V^{[N]}$.
This fiber loop matrix product state is shown in Fig.~\ref{fig:4}.

Our numerical construction of this fiber loop matrix product state starts from Eq.~\eqref{eq:BSE}.
We assume a $d$-dimensional Hilbert space for each mode of $U$.
Then the annihilation and creation operators $\hat{a}$, $\hat{b}$, $\hat{a}^{\dag}$, and $\hat{b}^{\dag}$ can each be written as a $d \times d$-dimensional matrix.
The $\hat{a}$/$\hat{a}^{\dag}$ and $\hat{b}$/$\hat{b}^{\dag}$ operators are represented by $d \times d$-dimensional matrices acting on different spaces, such that $U$ is a $d^{2} \times d^{2}$-dimensional matrix acting on the tensor product space, which can then be reshaped in a $d \times d \times d \times d$-dimensional tensor, see Eq.~\eqref{eq:EqV1}.
We always compute the matrix exponential for $U$ numerically and then reshape $U$ into a tensor of rank four, i.e.\ having four indices.
By contracting each rank-four tensor $U$ with its respective input state $|n_{i}\rangle$ we obtain the rank-three tensors $V^{[i]}$ of our fiber loop matrix product state.
Note that in this fiber loop matrix product state $\chi = d$.

\subsection{Entanglement and area law}
\label{sec:arealaw}

We now show that the MPS description of a time-bin single fiber loop is accurate for small $\chi$, i.e.\ the entanglement in the MPS description does not grow rapidly with system size.
More precisely, we now show that the entanglement entropy fulfills an area law (see Ref.~\cite{EiCrPl10} for a comprehensive presentation of the area law from the perspective of quantum many-body physics), i.e., is even independent of the system size.

\vspace{1mm}
\noindent \textbf{Theorem 1.}
For a train of $N$ input time-bin modes $\mathcal{T}_{1}$ to $\mathcal{T}_{N}$ in the multi-mode Fock state $|n_{1}\rangle \otimes |n_{2}\rangle \otimes \ldots \otimes |n_{N}\rangle$ acting on a single fiber loop, the entanglement entropy $E(i)$ of Eq.~\eqref{eq:E} for a bipartition between output time-bin modes $\tilde{\mathcal{T}}_{1}$ to $\tilde{\mathcal{T}}_{i}$ and $\tilde{\mathcal{T}}_{i+1}$ to $\tilde{\mathcal{T}}_{N}$ for any time-bin $i$ is upper-bounded by $g(n(i))$ where $n(i)$ denotes the average photon number inside the fiber loop in time-bin $i$ and
\begin{eqnarray}\label{eq:g}
g(n) & = & (1+n)\log_{2}(1+n) - n\log_{2}(n) .
\end{eqnarray}

\noindent \textbf{Proof.}
For every time-bin $i$, the state of the output fiber loop mode and the output time-bin modes $\tilde{\mathcal{T}}_{1}$ to $\tilde{\mathcal{T}}_{i}$ can be written as a Schmidt decomposition.
This can be achieved by writing $|\psi\rangle$ of Eq.~\eqref{eq:FS1} for $N = i$ and then computing the Schmidt decomposition (via standard tensor network techniques, see e.g.\ Refs.~\cite{VeMuCi08, Or14}) $|\psi\rangle = \sum_{k} \lambda^{[i]}_{k} |k\rangle_{\tilde{\mathcal{L}}_{i}} |k\rangle_{\tilde{\mathcal{T}}_{i} \otimes \tilde{\mathcal{T}}_{i-1} \otimes \ldots \otimes \tilde{\mathcal{T}}_{1}}$ with Schmidt coefficients $\lambda^{[i]}_{k}$.
The squared Schmidt coefficients are the eigenvalues of the reduced density matrix characterizing the fiber loop mode in time-bin $i$.
They allow us to compute the entanglement entropy $E(i)$ for the bipartition between the output fiber loop mode and the output time-bin modes $\tilde{\mathcal{T}}_{1}$ to $\tilde{\mathcal{T}}_{i}$ using Eq.~\eqref{eq:E}.
For any future time step $i + j$ where $j > 0$, the Schmidt coefficients $\lambda^{[i]}_{k}$ in $|\psi\rangle = \sum_{k} \lambda^{[i]}_{k} |k\rangle_{\tilde{\mathcal{L}}_{j} \otimes \tilde{\mathcal{T}}_{j} \otimes \ldots \otimes \tilde{\mathcal{T}}_{i+1}} |k\rangle_{\tilde{\mathcal{T}}_{i} \otimes \tilde{\mathcal{T}}_{i-1} \otimes \ldots \otimes \tilde{\mathcal{T}}_{1}}$ are equivalent to the previous Schmidt coefficients for time-bin $i$.
This is true because the input time-bin modes are a product state and the coupler is unitary.
So the entropy of the output fiber loop mode in time step $i$ is equivalent to the entropy of the output fiber loop mode together with output time-bin modes $\tilde{\mathcal{T}}_{i+1}$ to $\tilde{\mathcal{T}}_{i+j}$ in time step $i+j$.
Finally, we exploit the fact that among all states of the same average photon number the thermal state maximizes the entropy~\cite{SaNa17} (see also Ref.~\cite{WoGiCi06}) given by Eq.~\eqref{eq:g}. \rule{0.5em}{0.5em}

\vspace{1mm}
\noindent \textbf{Corollary 1.}
The entanglement entropy $E$ for any bipartition of the output state of a single fiber loop for an input multi-mode Fock state $|n_{1}\rangle \otimes |n_{2}\rangle \otimes \ldots \otimes |n_{N}\rangle$ satisfies an area law: $E \leq g(\max_{i}(n_{i}))$.
For an input state composed of the same number of photons $n$ in each input time-bin mode, this implies $E \leq g(n)$.

\vspace{1mm}
\noindent \textbf{Proof.}
Given $n(i)$ the mean photon number in the loop in time step $i$ and $n_{i+1}$ the input photon number in time-bin $i+1$, the average photon number in the fiber loop in time step $i+1$ is given by $n(i+1) = \cos^{2}(\theta)n(i)+\sin^{2}(\theta)n_{i+1}$.
This implies the bound $n(i+1) \leq \max(n(i), n_{i+1})$.
Thus, for the input multi-mode Fock state $|n_{1}\rangle \otimes n_{2}\rangle \otimes \ldots \otimes |n_{N}\rangle$ we obtain $n(i) \leq \max_{j \leq i}(n_{j})$.
This together with the previous theorem proves the area law for entanglement entropy $E \leq g(\max_{i}(n_{i}))$. \rule{0.5em}{0.5em}
\vspace{1mm}

Using the matrix product state representation of the single fiber loop setup, we have performed several numerical calculations that are presented in the following.
We programmed all required tensor network routines in the Tensor Network Theory Library~\cite{AlClJa17, TNTLibraryWebaddress}.
In all our numerical computations we assumed that every input time-bin mode $\mathcal{T}_{i}$ has the same photon number $n$, i.e.\ in the input multi-mode Fock state $|n_{1}\rangle \otimes |n_{2}\rangle \otimes \ldots \otimes |n_{N}\rangle$ we set $n_{i} = n$ for all $i$.
Furthermore, $\chi = d$ in all our computations, and all numerical results were converged with increasing values of $d$.

\begin{figure}
\centering
\includegraphics[width=0.9\linewidth]{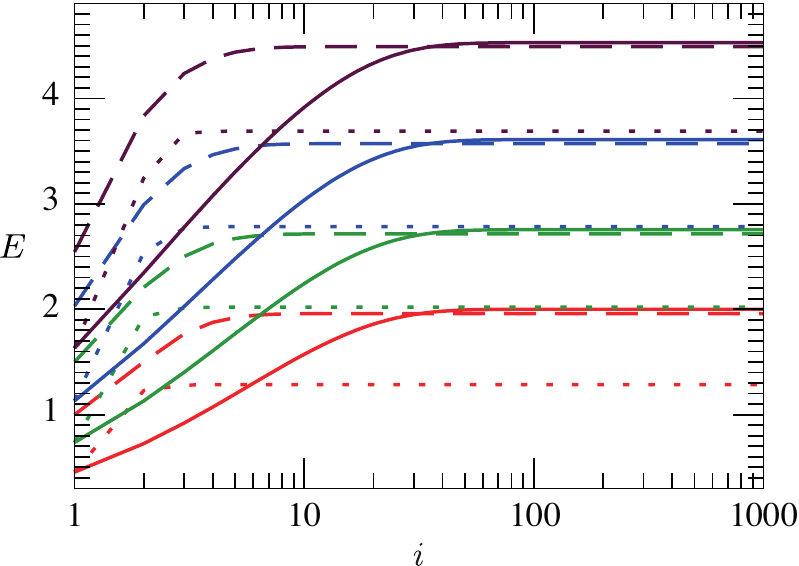}
\caption{\label{fig:5}
(Color online.)
Entanglement entropy $E$ as a function of time $i$ for $\theta = 0.4\pi$ (dotted), $0.25\pi$ (dashed), and $0.1\pi$ (solid), and $n = 1$ (red, lowest), $2$ (green, second-lowest), $4$ (blue, second-highest), and $8$ (purple, highest).
Our results are converged with increasing length $N$ and tensor dimensions $d$ and $\chi$ of the matrix product state: we used $N = 1100$ and $d = 81 = \chi$.
We computed the entanglement entropy via the canonical form of Eq.~\eqref{eq:canonical} and the formula of Eq.~\eqref{eq:E}.
}
\end{figure}

Figure~\ref{fig:5} shows the entanglement entropy $E(i)$ of the fiber loop mode in time-bin $i$ for different values of $n$ and $\theta$.
We observe that entanglement grows with time until it saturates at a final value that depends on $n$ and $\theta$.
For a fixed value of $\theta$, entanglement grows with increasing $n$.
For a fixed value of $n$, smaller $\theta$ lead to larger final amounts of entanglement but require longer saturation times.
Because $E(i)$ saturates to a constant value $E_{\mathrm{max}}$ for large values of $i$, i.e.\ after long enough time, our numerical results suggest that entanglement entropy has an upper bound independent of $N$ and thus satisfies an area law, as predicted in Theorem 1.
Entanglement saturation has an intuitive explanation: The entropy saturates when the loop reaches equilibrium, i.e., when the rate of photons going into the system is equal to the rate with which photons leak out to the detection mode.

\begin{figure}
\centering
\includegraphics[width=0.9\linewidth]{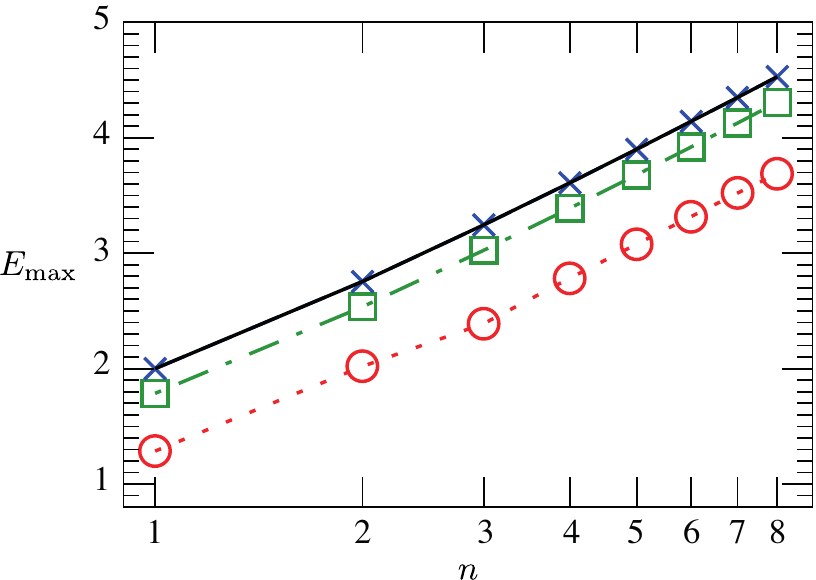}
\caption{\label{fig:6}
(Color online.)
Maximum entanglement entropy $E_{\mathrm{max}}$ as a function of $n$ for $\theta = 0.4\pi$ (circles, dotted), $\pi/3$ (squares, dash-dotted), $0.1\pi$ (crosses, dashed), and the analytical upper bound $(1+n)\log_{2}(1+n)-n\log_{2}(n)$ of Eq.~\eqref{eq:g} (solid).
Our results are converged with the same parameters as in Fig.~\ref{fig:5} and $E_{\mathrm{max}} = E(i=1000)$.
}
\end{figure}

Figure~\ref{fig:6} confirms the validity of our result on the area law upper bound for entanglement entropy.
We observe that the maximum entanglement entropy gets arbitrarily close to the analytical upper bound of Eq.~\eqref{eq:g} with decreasing value of $\theta$.
For each value of $\theta$ and the range of $n$ considered here, the dependence of $E_{\mathrm{max}}$ on $n$ is qualitatively well described by $(1+n)\log_{2}(1+n)-n\log_{2}(n)$.

\begin{figure}
\centering
\includegraphics[width=0.9\linewidth]{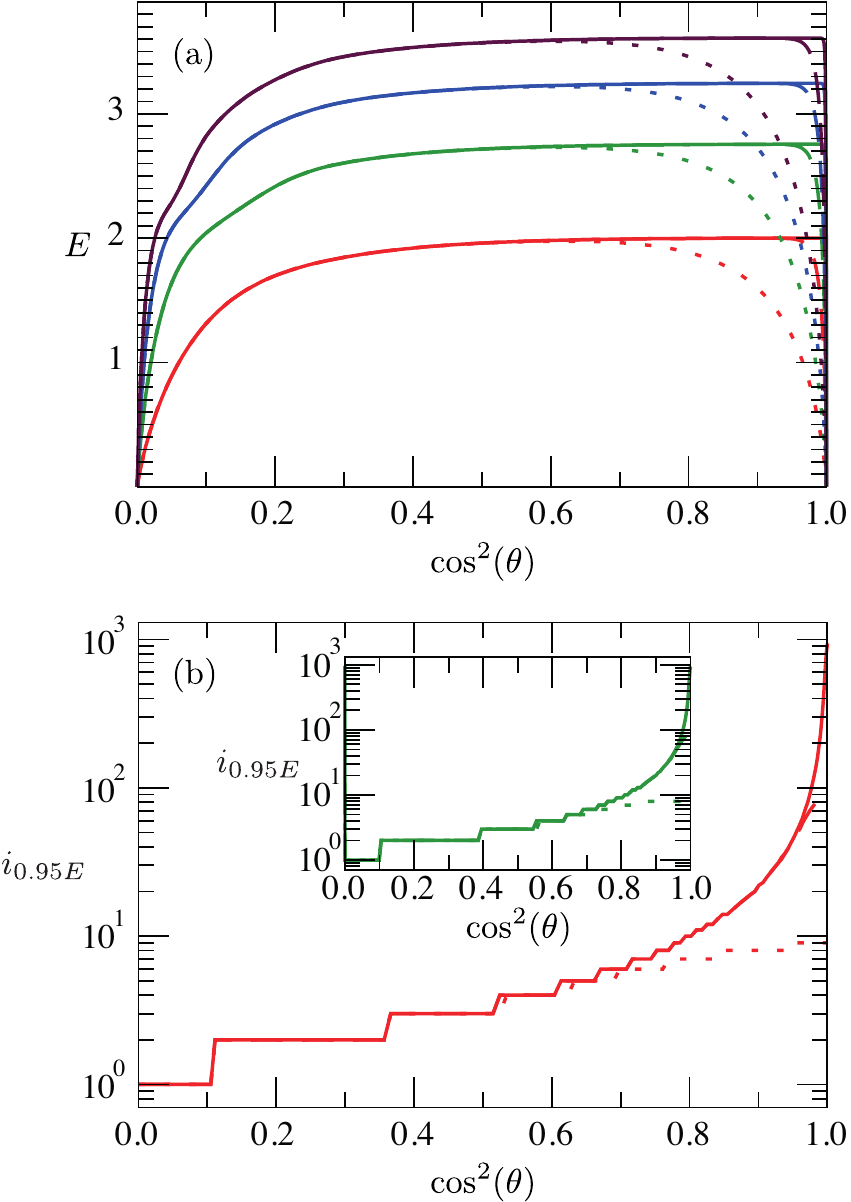}
\caption{\label{fig:7}
(Color online.)
(a) Entanglement entropy $E$ as a function of $\cos^{2}(\theta)$ for $i = 10$ (dotted), $100$ (dashed), $1000$ (solid), and $n = 1$ (red, lowest), $2$ (green, second-lowest), $3$ (blue, second-highest), and $4$ (purple, highest).
(b) Threshold time-bin $i_{0.95E}$ as a function of $\cos^{2}(\theta)$ for $i = 10$ (dotted), $100$ (dashed), $1000$ (solid), and $n = 1$ (main) and $2$ (inset).
The threshold time-bin is defined via $E(i_{0.95E}) > 0.95E(i)$ and $E(i_{0.95E}-1) < 0.95E(i)$: this quantifies the saturation time.
We have also analyzed $i_{0.95E}$ for $n = 3$ and $4$, and obtained very similar results to the cases $n = 1$ and $2$ shown here.
Our results are converged with $N$, $d$, and $\chi$.
We used $N = 1100$ and $d = 41 = \chi$.
}
\end{figure}

Figure~\ref{fig:7} shows the evolution of entanglement entropy over the full range of possible values for $\theta$.
We conclude that for every value of $n$, entanglement increases quickly with the transmission probability $\cos^{2}(\theta)$ and is already very close to the maximum possible $E$ when $\cos^{2}(\theta) > 0.4$.
The threshold time $i_{0.95E}$, which is defined in the caption of Fig.~\ref{fig:7}, seems to grow faster than exponentially with $\cos^{2}(\theta)$.
However, between $\cos^{2}(\theta) = 0.4$ and $\cos^{2}(\theta) = 0.7$ it takes on relatively small values $i_{0.95E} < 10$.
We observe that there exists an interesting trade-off -- depending on $\theta$ -- between the speed of convergence to the entanglement saturation value and how close this value is to the analytical upper bound of Eq.~\eqref{eq:g}.
On the one hand, a lower transmission probability leads to faster convergence to the entanglement saturation value -- as more photons are injected into the fiber loop in each time-bin -- but this saturation value is further away from the analytical upper bound.
On the other hand, a higher transmission probability leads to an entanglement saturation value that is the closer to the analytical upper bound but entanglement converges slower to this saturation value.

\subsection{Schmidt coefficients}

\begin{figure}
\centering
\includegraphics[width=0.9\linewidth]{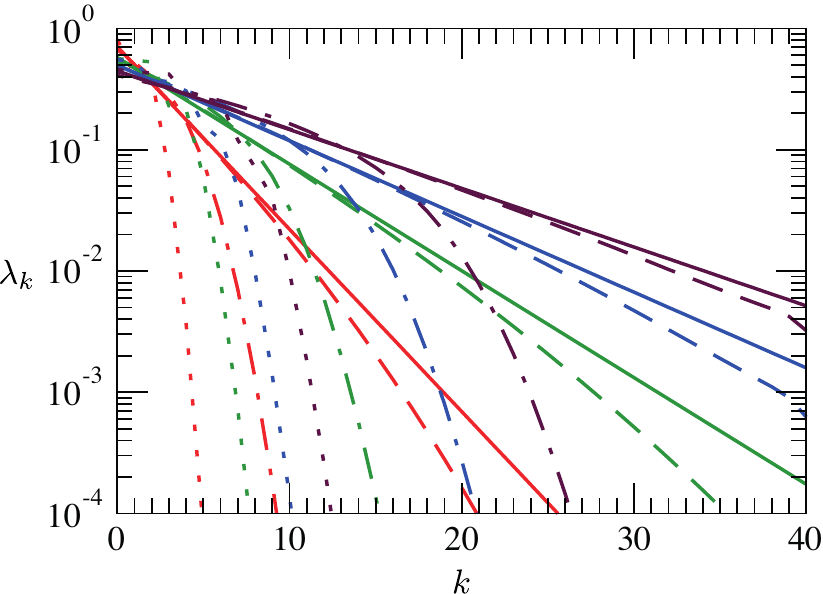}
\caption{\label{fig:8}
(Color online.)
Schmidt coefficients $\lambda_{k}$ for $\theta = 0.4\pi$ (dotted), $0.25\pi$ (dash-dotted), $0.1\pi$ (dashed), and analytical thermal state distribution $\sqrt{n^{k}/(n+1)^{k+1}}$ (solid), and $n = 1$ (red, lowest), $2$ (green, second-lowest), $3$ (blue, second-highest), and $4$ (purple, highest).
Our results are converged with $N$, $d$, $\chi$, and $i$.
We used $N = 1100$, $d = 41 = \chi$, and $i = 1000$.
}
\end{figure}

The remarkable fact that the entanglement saturation value converges to the analytical upper bound of Eq.~\eqref{eq:g} with decreasing value of $\theta$ suggests that the eigenvalues of the reduced density matrix of the fiber loop mode are close to those of the thermal state.
This is compared in Fig.~\ref{fig:8} where we plot the Schmidt coefficients for different input photon numbers $n$ and values of $\theta$ for a time-bin $i$ after entanglement saturation.
We conclude that for fixed $n$ the Schmidt values get closer to the thermal state distribution $\lambda_{k} = \sqrt{n^{k}/(n+1)^{k+1}}$ when the value of $\theta$ decreases.
All Schmidt coefficients decrease exponentially with $k$ but with a relatively small coefficient in the exponent.
Therefore relatively large values of $d$ and $\chi$ are needed for accurate matrix product state representations of these bosonic systems: for all problems considered in this article we observed convergence when $d = \chi \approx 10n$.

\subsection{Correlations}

\begin{figure}
\centering
\includegraphics[width=0.9\linewidth]{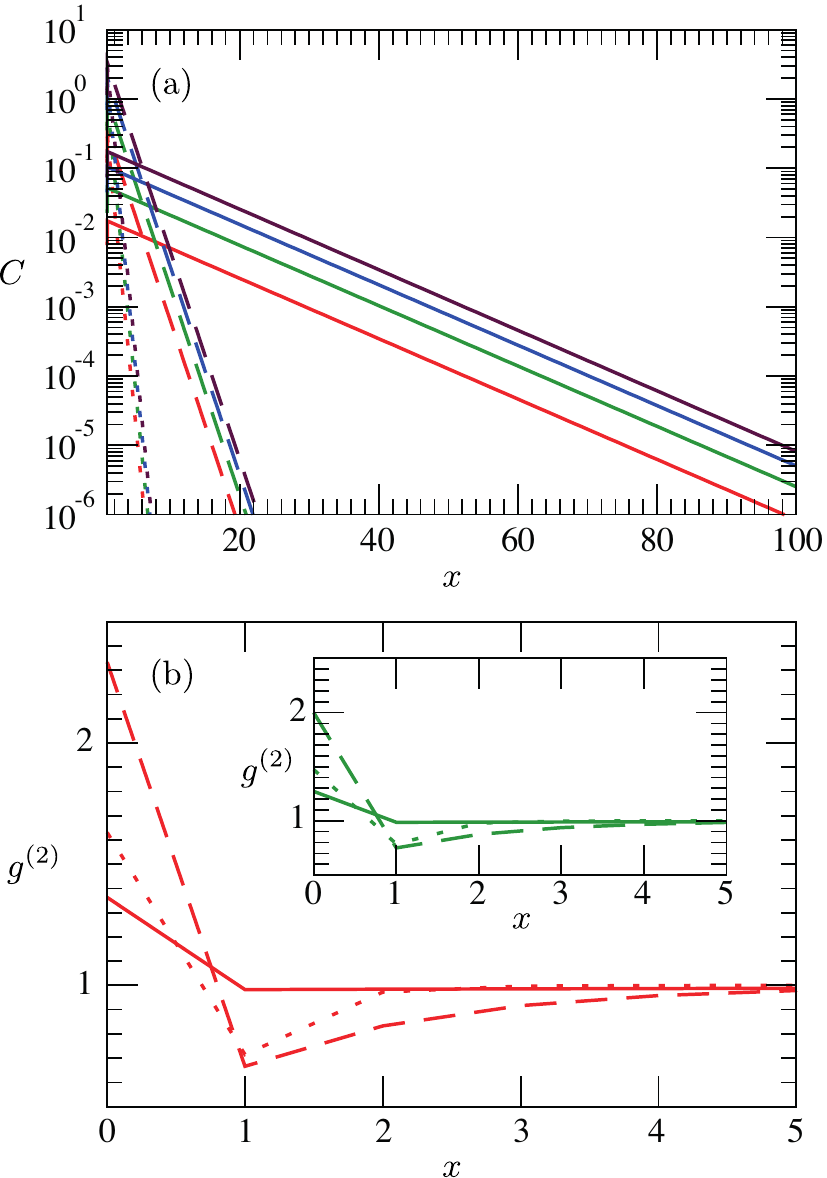}
\caption{\label{fig:9}
(Color online.)
(a) Correlation function $C(x) = |\langle \hat{n}_{i} \hat{n}_{i+x} \rangle - \langle \hat{n}_{i} \rangle \langle \hat{n}_{i+x} \rangle|$ for $\theta = 0.4\pi$ (dotted), $0.25\pi$ (dashed), $0.1\pi$ (solid), and $n = 1$ (red, lowest), $2$ (green, second-lowest), $3$ (blue, second-highest), and $4$ (purple, highest).
(b) Correlation function $g^{(2)}(x) = \langle \hat{n}_{i} \hat{n}_{i+x} \rangle / \langle \hat{n}_{i} \rangle^{2}$ for $\theta = 0.4\pi$ (dotted), $0.25\pi$ (dashed), $0.1\pi$ (solid), and $n = 1$ (main) and $2$ (inset).
We have also analyzed $g^{(2)}(x)$ for $n = 3$ and $4$, and obtained very similar results to the case $n = 2$ shown here.
Our results are converged with $N$, $d$, $\chi$, and $i$: we used $N = 1100$, $d = 41 = \chi$, and $i = 900$.
}
\end{figure}

In order to get a deeper understanding of the properties of a time-bin single fiber loop, we now analyze two-point correlation functions of the form $\langle \hat{n}_{i} \hat{n}_{i+x} \rangle$ where $\hat{n}_{i}$ and $\hat{n}_{i+x}$ act on the output time-bin modes $\tilde{\mathcal{T}}_{i}$ and $\tilde{\mathcal{T}}_{i+x}$.
Such expectation values can be computed exactly and efficiently with matrix product states, see e.g.\ Refs.~\cite{VeMuCi08, Or14}.
The required tensor contractions have a computational cost that scales like $\mathcal{O}(Nd\chi^{3})$.
Figure~\ref{fig:9} shows correlation functions computed after convergence with all matrix product state parameters.

We observe in Fig.~\ref{fig:9} (a) that the correlation function $C(x) = |\langle \hat{n}_{i} \hat{n}_{i+x} \rangle - \langle \hat{n}_{i} \rangle \langle \hat{n}_{i+x} \rangle|$ decreases exponentially with $x$, as expected from the matrix product state representation, see e.g.\ Refs.~\cite{VeMuCi08, Or14}.
This correlation function can be well approximated by $C(x) \approx C_{0} \exp(-x/\zeta)$ where the correlation length $\zeta$ depends only on $\theta$ -- and not on $n$ -- and the prefactor $C_{0}$ is determined by both $\theta$ and $n$.
For all our numerical results we have checked explicitly that $C(x)$ can be expressed in terms of the transmission probability $\cos^{2}(\theta)$ as $C(x) \approx \tilde{C}_{0}(\cos^{2}(\theta))^{x}$.
Hence for all our numerical results we obtain $\zeta^{-1} \approx -\log_{e}(\cos^{2}(\theta))$.

We also computed $C(x)$ using Eqs.~(A.4) and (A.6) from Ref.~\cite{WaEtAl16} which are closed-form expressions for $C(x) $ for interfering bosons and fully distinguishable particles (which exhibit no multi-particle interference), respectively.
We used the same parameters for this analytical calculation as for our numerical simulation.
We found for both distinguishable and indistinguishable photons that $C(x)$ shows exponential decay and that the decay constant is the same for both cases.
This shows that the dominant contribution to the exponential decrease of correlations is a classical effect, namely light leaking out of the loop mode over time.

It is important to point out that this does not mean that the considered state is classical, but rather that its quantumness is weak and not accessible through widely separated two-body correlations.
This is not surprising, as the entanglement between any bipartition is bounded by a small constant that is independent of the system size, i.e., entanglement satisfies an area law and is very localized.
Thus for long-ranged correlations the quantumness is quickly washed out.

Another means for analyzing the quantumness is given by the function $g^{(2)}(x)$ which is shown in Fig.~\ref{fig:9} (b).
We observe a signature of nonclassicality in short-ranged correlations $g^{(2)}(x=1) < 1$, which confirms the presence of entanglement in the system, but in a weak and short-ranged form.

\subsection{Sampling}

Simulating the photon counting statistics of a multi-mode quantum state of light is not an easy task.
The objective is to generate samples from the probability distribution $p(n_{1}, n_{2}, \ldots, n_{N})$ where the number of possible outcomes is $d^{N}$ and thus grows exponentially with the number of modes $N$.
This clearly is a daunting task that is, in full generality, classically intractable as shown by the original boson sampling proposal~\cite{AaAr11}.
However, when the quantum state of light is described by a matrix product state, then we can classically generate samples from the desired probability distribution efficiently.
This can be done following a sequential procedure that samples one random outcome per mode at a time and exploits the decomposition of a joint probability distribution as
\begin{eqnarray}
p(n_{1}, \ldots, n_{N}) & = & p(n_{N} | n_{N-1}, \ldots, n_{1}) \ldots p(n_{2} | n_{1}) p(n_{1}) . \quad\,\,\,\,\,
\end{eqnarray}

The first step of the algorithm calculates for every value of $n_{1}$ -- of which there exist $d = n_{\mathrm{max}} + 1$ values --  the probability of its outcome $p(n_{1}) = \langle \psi | ( |n_{1}\rangle \langle n_{1}| \otimes \mathcal{I}_{2, \ldots, N} ) |\psi \rangle$, where $|n_{1}\rangle \langle n_{1}|$ is the projector onto the photon number state $n_{1}$ of mode $1$ and $\mathcal{I}_{2, \ldots, N}$ is the identity operator on modes $2$ to $N$.
In order to compute $p(n_{1})$, we contract a tensor network that consists of the matrix product state representing the state as ket $|\psi\rangle$ with the bra $\langle \psi|$ interleaved with a tensor for the projector $|n_{1}\rangle \langle n_{1}|$.
After we have calculated all $d$ elements of $p(n_{1})$, we randomly select one $\tilde{n}_{1}$.
Then we update our state by generating $|\psi_{\tilde{n}_{1}}\rangle = \langle \tilde{n}_{1}|\psi \rangle$ where the bra $\langle \tilde{n}_{1}|$ acts only on mode $1$.
The result of this contraction is a new, unnormalized matrix product state $|\psi_{\tilde{n}_{1}}\rangle$ of size $N-1$.
Note that this new matrix product state satisfies the condition $\langle \psi_{\tilde{n}_{1}} | \psi_{\tilde{n}_{1}}\rangle = p(\tilde{n}_{1})$.

The second step now uses the state $|\psi_{\tilde{n}_{1}}\rangle$.
Firstly, we calculate the $d$ outcome probabilities $p(n_{2}, \tilde{n}_{1}) = \langle \psi_{\tilde{n}_{1}} | ( |n_{2}\rangle \langle n_{2}| \otimes \mathcal{I}_{3, \ldots, N} ) |\psi_{\tilde{n}_{1}} \rangle$ and randomly select a $\tilde{n}_{2}$ from the probability distribution $p(n_{2} | \tilde{n}_{1}) = p(n_{2}, \tilde{n}_{1}) / p(\tilde{n}_{1})$.
Secondly, we generate a new, unnormalized matrix product state $|\psi_{\tilde{n}_{1}, \tilde{n}_{2}}\rangle = \langle \tilde{n}_{2}|\psi_{\tilde{n}_{1}} \rangle$ of size $N-2$.
Continuing this procedure for the remaining $N-3$ output modes, we end up with one sample $(\tilde{n}_{1}, \tilde{n}_{2}, \ldots, \tilde{n}_{N})$ drawn according to the probability distribution $p(n_{1}, n_{2}, \ldots, n_{N})$.
Obviously we can generate as many samples as required by running this algorithm several times.
All tensor networks encountered in this algorithm can be contracted exactly and efficiently, and the overall computational cost scales as $\mathcal{O}(MN^{2}d^{2}\chi^{3})$ where $M$ denotes the number of samples.

An alternative simulation strategy exists for a single fiber loop.
As we can see in Fig.~\ref{fig:4}, the initial fiber loop is in a vacuum state and the first input time-bin corresponds to a photon number state.
Because the coupling preserves the total number of photons, measuring the output time-bin mode in the photon number basis collapses also the fiber loop into a specific photon number state.
This is easily generalized to other time steps, as we only need to adapt (and keep track of) the state of the fiber loop.

Although this alternative algorithm seems to be simpler, reformulating it for more complex circuits -- i.e.\ going beyond a single fiber loop -- quickly becomes a daunting task, as it is difficult to analytically derive the unitary for many couplings acting on many modes.
On the other hand, the extension of our matrix product state algorithm to more complex circuits is trivial, as all the complexity is moved to the numerics.
This is a big advantage of tensor networks and one of the reasons why we believe that they will become a useful tool for quantum optics.

\section{Route to complexity}
\label{sec:complexity}

The computational cost of contracting a tensor network depends on the number of closed loops in the graph of that tensor network, see e.g.\ Refs.~\cite{VeMuCi08, CiVe09, Or14}.
Here the notion of a loop is synonymous to the notion of a simple cycle in graph theory~\cite{Di05}.\footnote{
A simple cycle is a path of vertices and edges in which each vertex can be reached from itself via a sequence of vertices that starts and ends at the same vertex and has no repetitions of vertices in between.}
So if we want to increase the computational complexity of a quantum optical architecture, we need to generate a tensor network state with increasing number of loops.
As we will explain in detail below, this can be achieved by increasing the number of interferometric loops, where these loops might have a physical existence in the experimental implementation or result from the interference between different time-bin modes.
We will present several experimental proposals based on time-bin fiber loops and discuss their computational complexity with the help of their tensor network representation.

\subsection{Tensor network states with multiple loops}
\label{sec:lattice}

We have seen in Sec.~\ref{sec:classicalsimu} that a single fiber loop can be represented by a matrix product state, which is a tensor network without cycles, see Fig.~\ref{fig:4}.
In the following we discuss several approaches to increase the computational complexity of the generated tensor network state by adding more interferometric loops to the circuit.

Besides matrix product states, there exist many other interesting types of tensor networks, and some of the most widely used tensor network types are explained in Refs.~\cite{VeMuCi08, CiVe09, Or14}.
Of particular importance for this article are projected entangled pair states~\cite{VeCi04, MuVeCi07} which represent a very natural generalization of matrix product states to higher dimensions.
Projected entangled pair states can contain much more entanglement than matrix product states but their exact tensor network contraction is known to be computationally hard~\cite{ScWoVeCi07}.
However, there exist efficient methods for approximate tensor network contractions in the context of projected entangled pair states, see e.g.\ Refs.~\cite{VeCi04, MuVeCi07, VeMuCi08, Or14, LuCiBa14NJP, LuCiBa14PRB}.
Here we will see how projected entangled pair states can be generated with fiber loops.

\subsubsection{Increasing the ancillary system dimension}

One example of more complex tensor network states is given by matrix product-like states that have a bond dimension that does not satisfy an area law.
As the bond dimension in a matrix product state is related to the dimension of the ancillary system in the sequential generation of this matrix product state -- see discussion in Sec.~\ref{sec:MPSanalysis} and Refs.~\cite{ScSoVeCiWo05, ScHaWoCiSo07} -- we can increase the computational complexity by enlarging the degrees of freedom of the ancillary system.
This can be achieved, e.g., by increasing the number of interconnected fiber loops that interact with the incoming time-bin photons.

\begin{figure}
\centering
\includegraphics[width=0.9\linewidth]{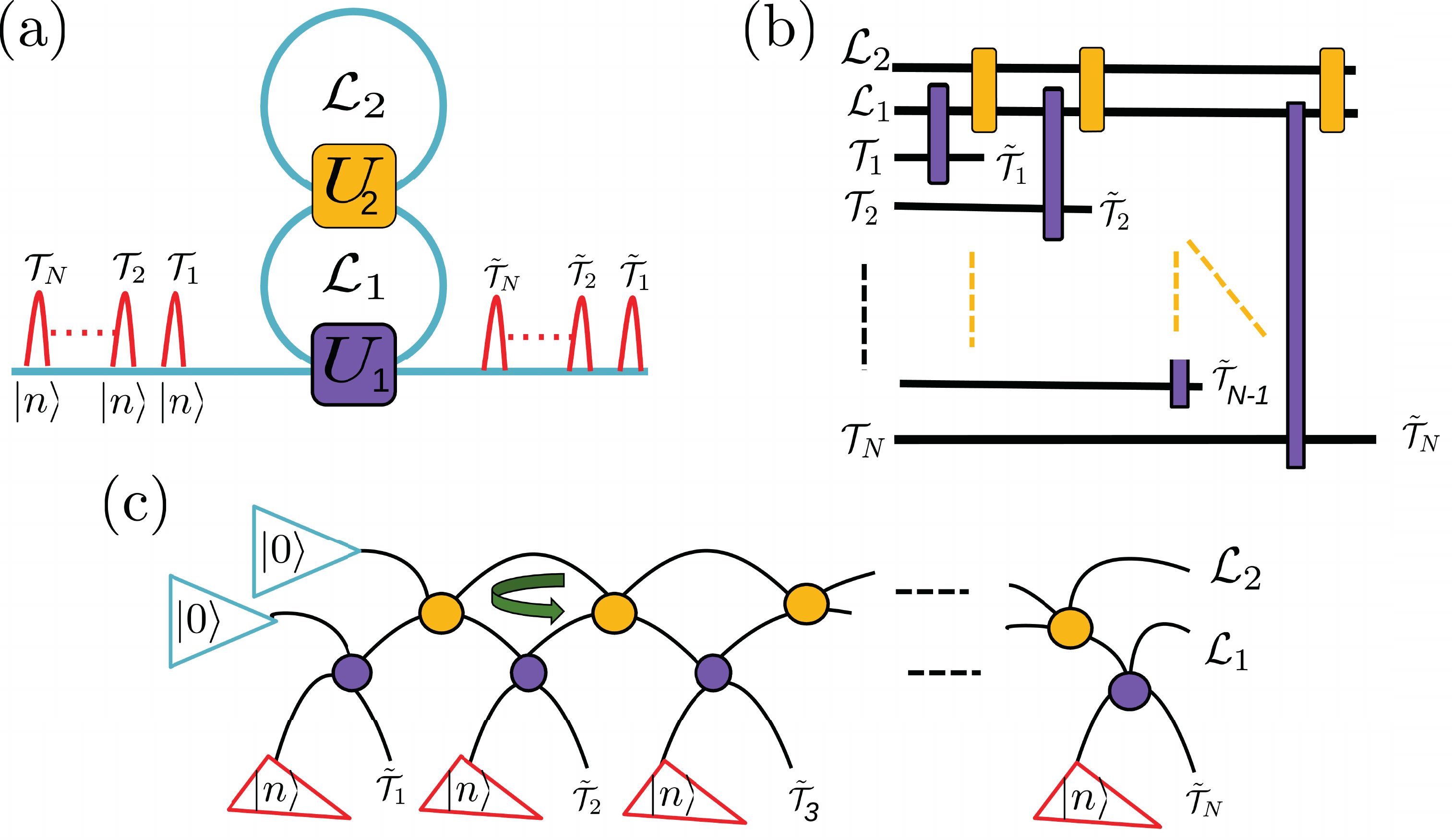}
\caption{\label{fig:10}
(Color online.)
(a) A tower of two ancillary fiber loops $\mathcal{L}_{1}$ and $\mathcal{L}_{2}$ is interacting with a train of input time-bin modes $\mathcal{T}_{i}$ of photon numbers $n_{i}$ and generating a train of output time-bin modes $\tilde{\mathcal{T}}_{i}$.
(b) The circuit is composed of a ladder of identical coupler gates realizing the unitary $\hat{U}_{1}$ (purple, dark box) between the input time-bin modes $\mathcal{T}_{i}$ and $\mathcal{L}_{i}$ at time step $i$.
Note that, different from Fig.~\ref{fig:2}, now the fiber loop system $\mathcal{L}$ does not move one step down the ladder after every coupling.
Each coupling $\hat{U}_{1}$ is followed by a coupler $\hat{U}_{2}$ (yellow, light box) between the two fiber loop modes $\mathcal{L}_{1}$ and $\mathcal{L}_{2}$.
(c) The corresponding tensor network state is obtained by replacing the coupling gates $\hat{U}_{1}$ (purple, dark box) and $\hat{U}_{2}$ (yellow, light box) at each time step $i$ by their corresponding graph vertices.
The input states of the loop and time-bin modes are the blue (left) and red (bottom) triangles, respectively, which represent tensors of rank one.
The green arrow highlights a closed loop in the tensor network state.
}
\end{figure}

Figure~\ref{fig:10} shows an example of an ancillary structure composed of two fiber loops.
A detailed analysis of the coupling between the input time-bin modes $\mathcal{T}_{i}$ with both loop modes $\mathcal{L}_{1}$ and $\mathcal{L}_{2}$, as sketched in Fig.~\ref{fig:10} (a), leads to the circuit diagram shown in Fig.~\ref{fig:10} (b).
The beam splitter $\hat{U}_{1}$ between the first loop $\mathcal{L}_{1}$ and the input time-bin mode $\mathcal{T}_{i}$ at time step $i$ is followed by a beam splitter $\hat{U}_{2}$ between the two loop modes.
It is straightforward to obtain the tensor network graph for the output state from the circuit diagram: we just need to identify the initial states and gates as the vertices in the tensor network state.
We can see in Fig.~\ref{fig:10} (c) that the double layer of ancillary loop modes generates closed loops in the tensor network representation.
Clearly we could go beyond this and generate even more loops in the tensor network graph and thus even more complex tensor network states, simply by including more fiber loops in the ancillary system.
The resulting tensor network state would resemble the tensor network graph that is obtained for the norm computation of a projected entangled pair state~\cite{VeCi04, MuVeCi07} and its exact contraction would be computationally hard~\cite{ScWoVeCi07}.

\subsubsection{Concatenating fiber loops}

\begin{figure}
\centering
\includegraphics[width=0.75\linewidth]{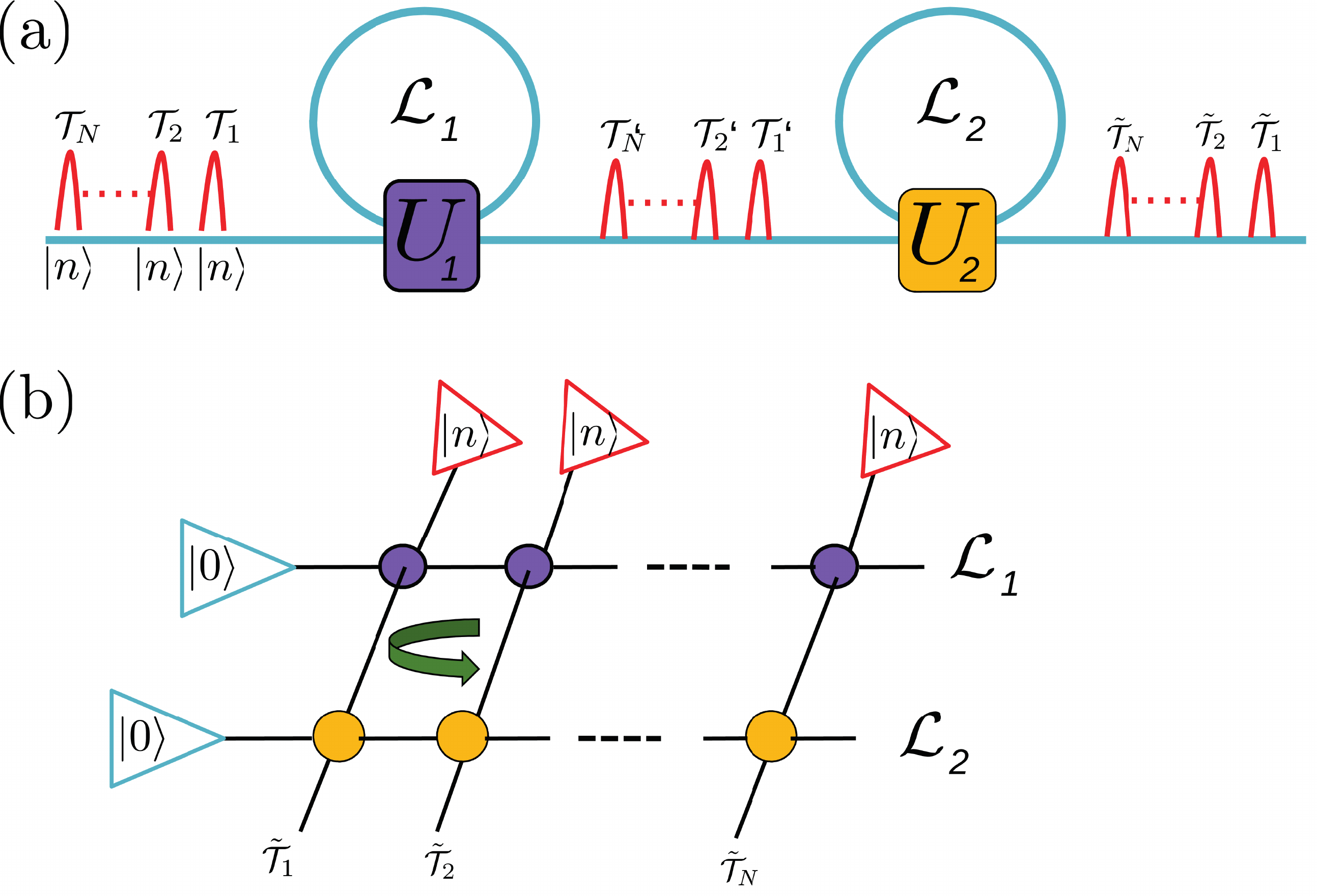}
\caption{\label{fig:11}
(Color online.)
(a) Scheme of two concatenated fiber loops $\mathcal{L}_{1}$ and $\mathcal{L}_{2}$, where $\mathcal{T}_{i}$ and $\tilde{\mathcal{T}}_{i}$ represent input and output time-bin modes, respectively, and $\mathcal{T}_{i}'$ represent the time-bin modes in the middle of the interferometer.
$\hat{U}_{1}$ and $\hat{U}_{2}$ are beam splitters.
(b) Tensor network representation of the output state for $N$ time-bin modes after passing through the two fiber loops.
The purple (dark)/yellow (light) circles represent tensors with two input and two output indices where each index corresponds to either an input or output degree of freedom of the coupling $\hat{U}_{1}$/$\hat{U}_{2}$.
The blue (left) and red (top) triangles represent tensors with one index and this index corresponds to the input state of the loop and time-bin modes, respectively.
}
\end{figure}

An alternative approach for the generation of tensor network states with cycles is given by concatenating fiber loops.
In Fig.~\ref{fig:4} we can see that the horizontal edges of the tensor network graph correspond to the physical indices of the fiber loop mode $\mathcal{L}$ at different time steps $i$ and the vertical edges correspond to the input and output time-bin modes $\mathcal{T}_{i}$ and $\tilde{\mathcal{T}}_{i}$, respectively, where the latter are open, i.e.\ not contracted, indices as they correspond to the physical indices of the final state.
When we concatenate two fiber loops, as depicted in Fig.~\ref{fig:11} (a), the output time-bin modes of the first loop $\mathcal{L}_{1}$ become the input modes of the second fiber loop $\mathcal{L}_{2}$.
This creates a tensor network graph with an additional horizontal line, shown in Fig.~\ref{fig:11} (b), that encodes the evolution of the internal degree of freedom of the second fiber loop $\mathcal{L}_{2}$ over time.
Note that because the couplings $\hat{U}_{1}$ and $\hat{U}_{2}$ are time-invariant, all the tensors in the first and second horizontal line are the same, respectively.
Obviously we could systematically increase the complexity by concatenating more and more fiber loops.
Just as in the previous scenario, this procedure would create a tensor network state similar to the norm tensor network of a projected entangled pair state~\cite{VeCi04, MuVeCi07} which is hard to contract exactly~\cite{ScWoVeCi07}.

\subsubsection{Multiple parallel time-bin chains}

\begin{figure}
\centering
\includegraphics[width=0.9\linewidth]{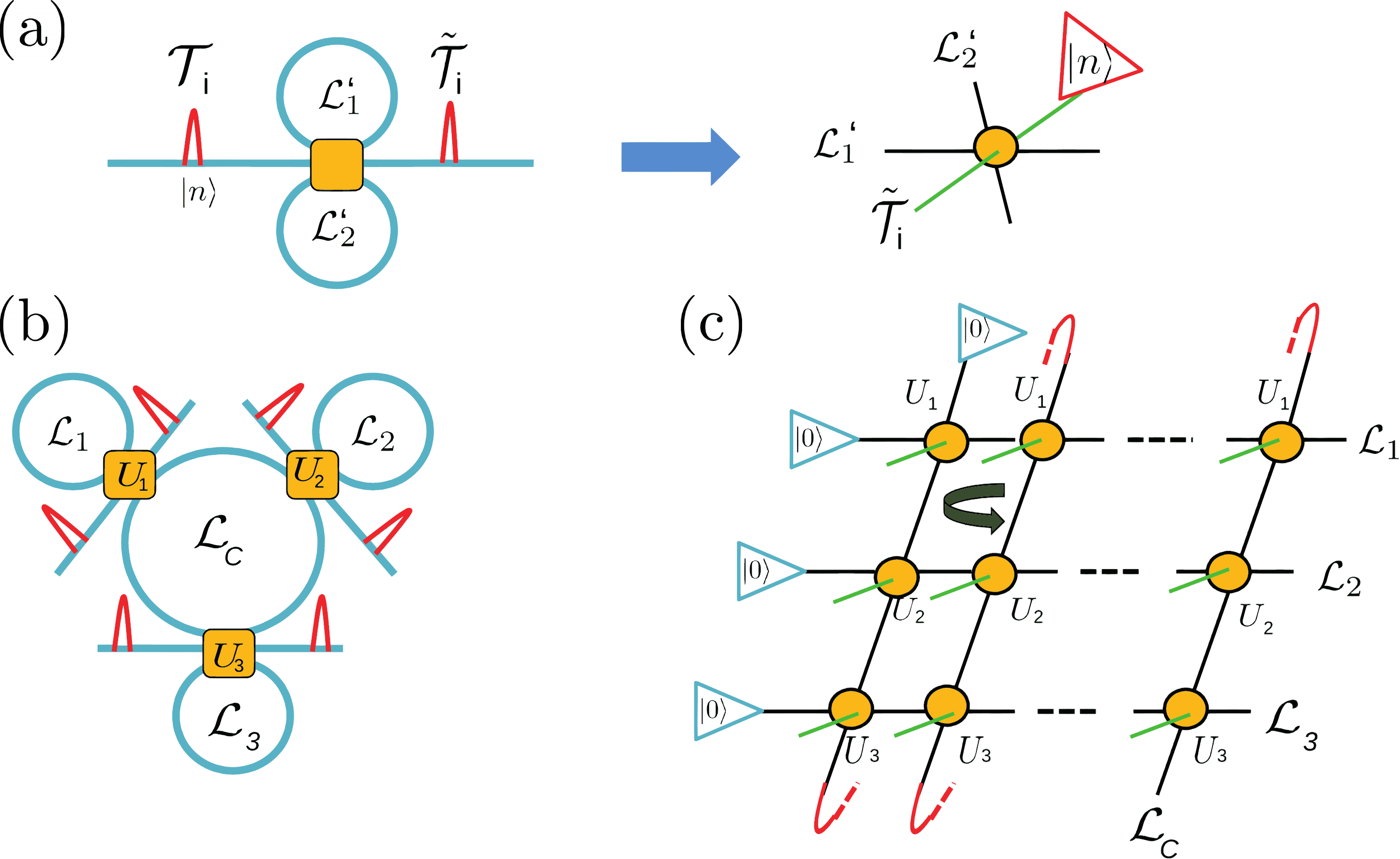}
\caption{\label{fig:12}
(Color online.)
(a) A tritter is defined by a unitary with three inputs and three outputs.
For a fixed time step $i$ and input time-bin $\mathcal{T}_{i}$, after contraction of the input state $|n\rangle$ (red triangle) with the tritter vertex, we obtain a tensor with five indices where four indices correspond to the two initial and final fiber loop modes and the remaining index corresponds to the output time-bin mode $\tilde{\mathcal{T}}_{i}$ (green, diagonal edge).
(b) Three tritters have a local fiber loop each, $\mathcal{L}_{1}$, $\mathcal{L}_{2}$, and $\mathcal{L}_{3}$, and a commonly shared fiber loop $\mathcal{L}_{\mathrm{C}}$.
(c) This generates a projected entangled pair state with cylindrical boundary conditions.
The cylindrical boundary conditions result from the fact that every final vertical edge of a given column of the graph, leaving a $\hat{U}_{3}$ vertex, becomes the uppermost vertical input edge, entering a $\hat{U}_{1}$ vertex of the following column.
}
\end{figure}

The tensor network formalism that we have just described for the tower and concatenation of fiber loops can be extended to more general interferometric schemes, composed of any number of fiber loops or time-bin modes in arbitrary architectures.
Additionally, we could also have multiple input and output trains of time-bin pulses that interact with each other simultaneously through a complex network of fiber loops.
As an example, Fig.~\ref{fig:12} shows how tritters, i.e.\ three-mode interferometers, can be used to generate a very complex tensor network, namely a projected entangled pair state with cylindrical boundary conditions.

It is also not difficult to see that one can alternatively generate a tensor network with cylindrical boundary conditions using a tower of two loops, as in Fig.~\ref{fig:10}, by choosing the first loop $\mathcal{L}_{1}$ to be larger than the second loop $\mathcal{L}_{2}$.

\subsection{Hard instances}
\label{sec:hard}

The intuition that the computational complexity of a tensor network contraction increases with the number of simple cycles contained in the tensor network graph can be formulated mathematically via the concept of the treewidth of a graph.
This is a computable quantity that quantifies how far a graph is from being acyclic~\cite{MaSh08}.

\subsubsection{Treewidth of a graph}

\begin{figure}
\includegraphics[width=0.7\linewidth]{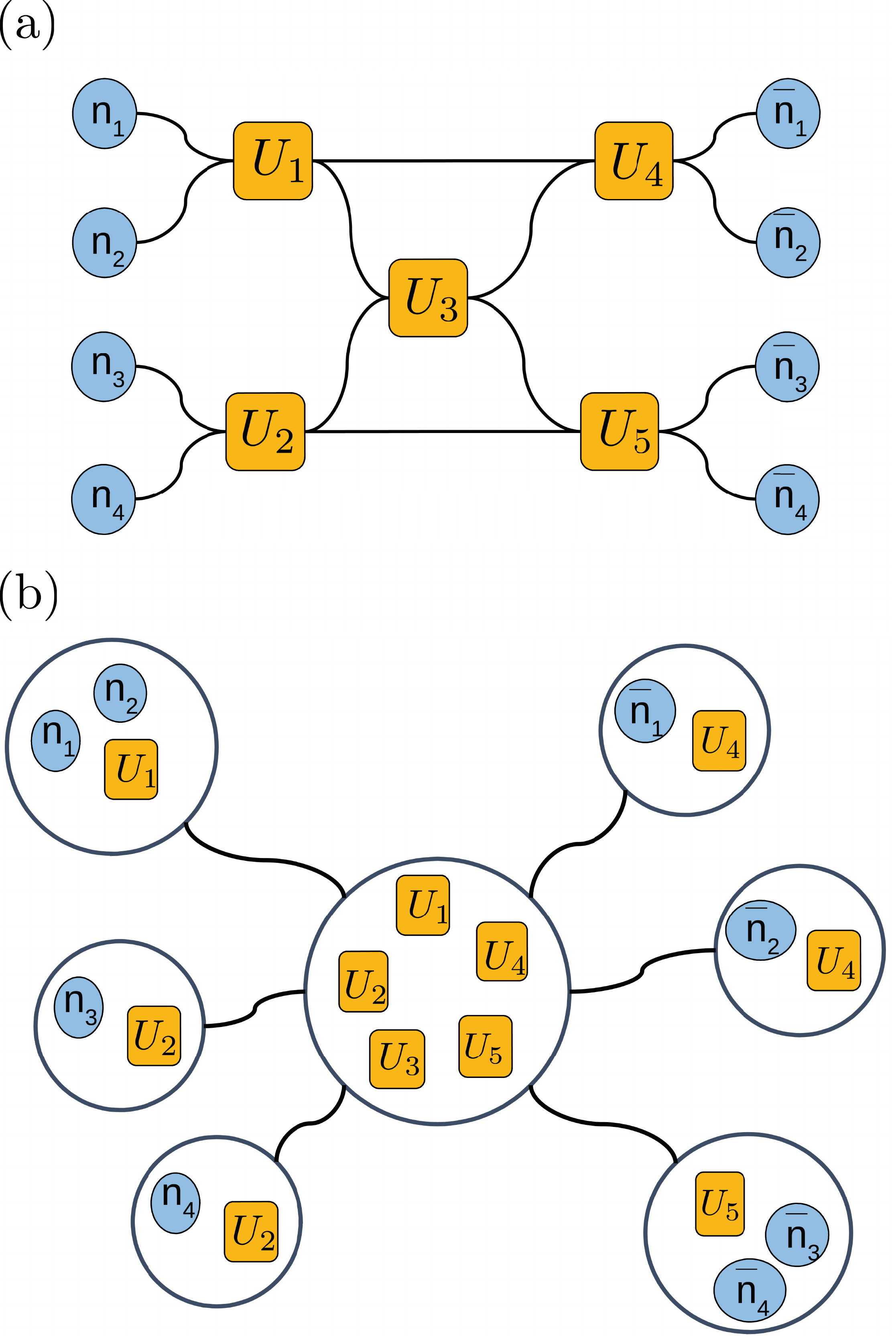}
\caption{\label{fig:13}
(Color online.)
(a) This tensor network $G$ represents the evolution of four initial states with photon numbers $n_{1}$ to $n_{4}$ over a set of five couplers $U_{1}$ to $U_{5}$ and leads to measurement outcomes of photon numbers $\bar{n}_{1}$ to $\bar{n}_{4}$.
(b) A tree decomposition results from creating bags that contain tensors of $G$ following the three rules explained in the main text just above Eq.~\eqref{eq:width}.
The width of this decomposition is four, as the largest bag -- namely the central one -- contains five tensors.
}
\end{figure}

A tree graph, or acyclic graph, is a graph where every path between two vertices is unique, see Fig.~\ref{fig:13} (b) for an example.
As shown in Ref.~\cite{MaSh08} a tree tensor network can be contracted efficiently, where the running time scales polynomially with the number of vertices and the bond dimension.
Note that a tree tensor network graph is more general than a one-dimensional chain, and thus a matrix product state, as it may contain attached branches.

For a cyclic graph, such as the one in Fig.~\ref{fig:13} (a), the notion of tree decomposition is an ensemble of operations that maps every vertex $v$ of a tree graph $G_{\mathrm{tree}}$ into a subset $B(v)$ of vertices of the initial cyclic graph $G$, called bags.
The mapping must satisfy the following three rules:
\begin{itemize}
\item Each vertex of $G$ must appear in at least one bag.
\item For each edge of $G$ at least one bag must contain both of its end vertices.
\item For a every vertex $v$ of $G$ the set of vertices of $G_{\mathrm{tree}}$ such that their corresponding bags contain $v$ form a connected subtree.
\end{itemize}
The width of a given tree decomposition is defined as
\begin{eqnarray}\label{eq:width}
\mathrm{width}(G_{\mathrm{tree}}) & = & \max_{v \in V(G_{\mathrm{tree}})}|B(v)|-1
\end{eqnarray}
where $V(G_{\mathrm{tree}})$ is the set of vertices of $G_{\mathrm{tree}}$ and $|B(v)|$ is the size of the bag corresponding to vertex $v$ of $G_{\mathrm{tree}}$, see Ref.~\cite{MaSh08} for details.
For a specific tree decomposition, it is easy to calculate its width.
But a given cyclic graph can have different decompositions of different widths.
Therefore, we define the treewidth $\mathrm{tw}(G)$ as the minimum width over all the possible tree decompositions of $G$, and this quantifies how far the graph is from a tree.
The exact calculation of the treewidth of a general tensor network -- which is equivalent to finding the optimal tensor contraction order -- is a NP-complete problem~\cite{Bo86, Bo05}.

\subsubsection{Sufficient condition for efficient simulation}

If we want to adapt the main result of Ref.~\cite{MaSh08} to the harmonic oscillator setup of quantum optics, then we need to take the possible bunching of photons into account.
This can be achieved by defining the local dimension $d$ of every tensor to be as large as the total number of photons $M$ plus one, $d = M+1$.
The next lemma follows immediately.

\vspace{1mm}
\noindent \textbf{Lemma 1.}
Let $G_{A}$ denote the tensor network graph for a photonic architecture $A$ of $N$ modes containing a maximum total number of $M$ photons and $T$ couplers.
Then the output probability distribution of local measurements can be deterministically computed in $T^{\mathcal{O}(1)}M^{\mathcal{O}(\mathrm{tw}(G_{A}))}$ time.
\vspace{1mm}

As opposed to the tensor contraction of finite-dimensional systems (qudits, spins) where a treewidth scaling logarithmically with the size of the system leads to a $\mathrm{poly}(\mathrm{tw}(G))$ time algorithm, here we obtain a quasi-polynomial time algorithm for logarithmically scaling treewidth.
We could restore the polynomial scaling occuring in finite-dimensional systems if we could truncate the physical dimension $d$ to a constant value, e.g.\ to a few times the average photon number on each mode, an efficiently computable quantity.
Unfortunately, it is non-trivial to prove that such an approach leads to an algorithm that can approximate the tensor network contraction with error $\epsilon$ and has a running time that scales polynomially both in $1/\epsilon$ and the size of the system, $M$ and $N$.
This proof is a task that we leave for future work.

Although the calculation of the treewidth is NP-complete, finding an upper bound that is sub-logarithmic is sufficient to prove that a tensor network has an efficient contraction, and that is a simpler task than finding the optimal contraction.
In order to prove the hardness of a tensor contraction one needs to show that the treewidth has a super-logarithmical lower bound, which for some classes of graphs is easy to do.
E.g.\ a lattice graph $G$ with $|V(G)|$ vertices has a treewidth scaling as $\sqrt{|V(G)|}$.

\subsubsection{A route to resilient quantum advantage architectures}

One of the most pursued routes to quantum advantage is based on searching for non-universal and easy to implement quantum architectures that generate sampling distributions that are hard to reproduce by a classical computer.
The main path to prove hardness of sampling in a given quantum architecture is to prove that at least one probability outcome is \#P-hard to compute (for exact sampling) or most instances are \#P-hard to compute (for approximate sampling).
Because the hardest instances of tensor networks are \#P-hard to contract~\cite{ScWoVeCi07}, it is tempting to leverage all knowledge accumulated about tensor networks in order to search for good candidates of a quantum advantage architecture.
In this direction the previous theorem is a very useful tool that allows us to discard some photonic architectures that are classically simulatable.
At the same time this theorem provides hints of where to look for good architectures that are computationally hard for sampling but less demanding in the number of resources and more resilient to imperfections than current proposals.
E.g.\ we might be able to trade off depth of the circuit of current boson sampling proposals, which usually results in added losses in the real experiment, by moving from a planar circuit architecture to a three-dimensional structure that generates tensor networks of similar treewidths than the corresponding planar circuit but in a more compact volume and therefore reduces the losses and the noise.
We leave further exploration of this for future work.

\section{Conclusions and outlook}

In this article we formulated time-bin linear quantum optics in the framework of tensor network states.
This allowed us to make a first step towards using tensor network states for the simulation and design of large quantum optical experiments.

First we focussed on a single fiber loop and demonstrated that it has an efficient matrix product state representation.
We extensively investigated quantities natural in many-body physics, such as the entanglement entropy and the correlations in the system.
We proved an area law for the single fiber loop architecture and showed numerically that our area law upper bound is tight and experimentally achievable.
We also observed an exponential decay of correlations and of Schmidt coefficients, which together with the area law, are consistent with a matrix product state representation of finite bond dimension that is classically efficiently simulatable.

We discussed the connection between interferometric loops and cycles in the tensor network representation, known to be the source of increasing computational complexity within the tensor network framework.
Then we showed how to build more complex setups composed of several fiber loops and tritters, and we demonstrated that these experimental setups can generate tensor networks, such as projected entangled pair states, whose contraction is known to be hard.

Boson sampling is a quantum optical proposal for demonstrating a quantum advantage and has attracted a lot of interest from both theoreticians and experimentalists.
As in other quantum advantage proposals, the effects of experimental imperfections, such as photon losses, photon distinguishability, and general noise, will play a decisive role in an experimental realization.
Therefore, identifying novel photonic architectures that are more resilient to those experimental imperfections will be crucial.
Because experimental imperfections can be easily included in tensor network simulations, we believe that tensor networks have the potential to guide the design of future boson sampling experiments.

An interesting next step is the development of novel numerical tensor network techniques tailored to the unbounded nature of the quantum optical scenario.
We saw in our simulations that this unbounded nature of the optical modes adds an additional degree of complexity:
In order to obtain accurate numerical results, we needed to set the physical dimension of the system to be one order of magnitude larger than the average number of photons per mode.
Interestingly, we also observed in our quantum optical simulations that bond and physical dimension could be treated in the same way.
A possible approach, inspired by the similarities between bond and physical dimension, would be the development of approximation algorithms that truncate both bond and physical dimensions during the tensor network contraction.
There exist already highly efficient algorithms that can compute approximations of projected entangled pair state contractions and provide estimates of the approximation errors, see e.g.\ Refs.~\cite{LuCiBa14NJP, LuCiBa14PRB} for recent algorithmic proposals.
It is tempting to apply these methods -- and modified versions that truncate both bond and physical dimensions -- to classically simulate the boson sampling experiments proposed in Sec.~\ref{sec:complexity}.
This would allow us to refine the boundary between boson sampling experiments that can and cannot be currently classically simulated, using state-of-the-art tensor network methods.

During the completion of this article we learned about Ref.~\cite{DhEnSaBaSiPl17} that also discusses the use of tensor networks to simulate quantum optical experiments.

\section{Acknowledgements}

M.L. and D.J. acknowledge funding from the Networked Quantum Information Technologies Hub (NQIT) of the UK National Quantum Technology Programme as well as from the EPSRC grant ``Tensor Network Theory for strongly correlated quantum systems'' (EP/K038311/1).
J.J.R. acknowledges NWO Rubicon.
M.S.K. acknowledges the Royal Society and the Samsung GRO grant.
I.W. acknowledges ERC Advance Grant MOQUACINO.
R.G.-P. is Research Associate of the F.R.S.-FNRS.
R.G.-P. and I. W. acknowledge funding from the Fondation Wiener-Anspach.
A.A.V., M.S.K., and I.W. acknowledge support from Physical Science Research Council (project EP/K034480/1)
A.A.V. thanks Germ\'{a}n Sierra and Juan Jos\'{e} Garc\'{i}a-Ripoll for helpful discussions.
We are also grateful to Joelle Boutari for helpful discussions.

\end{document}